\documentclass{iau-JDSS}
\usepackage{graphicx}

\title[Stability of pulsar rotational and orbital periods] 
{Stability of pulsar rotational and orbital periods}

\author[Sergei Kopeikin]   
{Sergei Kopeikin}

\affiliation{Department of Physics and Astronomy, University of Missouri,
Columbia, MO 65211, USA \break email: kopeikins@missouri.edu}

\pubyear{2010}
\volume{Volume 15}  
\pagerange{???? - ????}
\date{September 15, 2009 and in revised form ??}
\jname{Highlights of Astronomy}
\editors{Karel van der Hucht, ed.}
\begin{document}

\maketitle

\begin{abstract}
Millisecond and binary pulsars are the most stable astronomical standards of frequency. They can be applied to solving a number of problems in astronomy and time-keeping metrology including the search for a stochastic gravitational wave background in the early universe, testing general relativity, and establishing a new time-scale. The full exploration of pulsar properties requires that proper unbiased estimates of spin and orbital parameters of the pulsar be obtained. These estimates depend essentially on the random noise components present in pulsar timing residuals. The instrumental white noise has predictable statistical properties and makes no harm for interpretation of timing observations, while the astrophysical/geophysical low-frequency noise corrupts them, thus, reducing the quality of tests of general relativity and decreasing the stability of the pulsar time scale.
\end{abstract}
Timing observations of single and, especially, binary millisecond pulsars are widely recognized as extremely important for progression in a number of branches of modern astronomy and time-keeping metrology. In particular, the implication of pulsar timing for testing general relativity in the strong-field regime [\cite{2009CQGra..26g3001K}] and setting up the upper limit on the energy density of the stochastic gravitational wave background in the early universe [\cite{2009PASA...26..103H}] are among the most stimulating on-going research activities. Another important aspect of pulsar timing is metrology of time on long time intervals providing a time-scale based on the high-stable rotation of millisecond pulsars around its own axis [\cite{ilyasovpt,2007HiA....14..479I}] and/or center of mass of a binary system [\cite{petit,1998AstL...24..228I}].

The accuracy of pulsar timing observations is now approaching 10 nanoseconds. Such high precision requires construction of adequate data processing software taking proper account of the relevant physical effects that can contribute to the variations of timing residuals [\cite{2006MNRAS.372.1549E}]. Usually, the procedure for estimating pulsar parameters is based on the premise that white noise is a dominant source of instability in times of pulse's arrival. However, long-term monitoring of pulsars certainly reveals the presence of a non-white component of the noise having the external origin [\cite{2006ChJAS...6b.169H}]. This noise is called {\it red} as it has a spectrum that diverges at zero frequency. The lower the timing activity of the pulsar, the further toward low frequencies one must look in order to detect the {\it red} noise in the timing residuals. The {\it red} noise also makes the residuals correlated on long time intervals.

Developing a computer code accounting for effects of the {\it red} noise on timing residuals and estimates of parameters is worthwhile. It requires further improving the model of the {\it red} noise \cite{1997MNRAS.288..129K} and the methods of estimation of the stability of the pulsar rotational/orbital phase [\cite{1999MNRAS.305..563K,2004MNRAS.355..395K}]. The present state-of-the art statistical analysis of pulsar timing data in the presence of a {\it red} noise has not yet reached the required level of completeness and a more elaborate technique has to be invented.

Analysis of the stability of pulsar's rotational and orbital phase is more informative in time rather than in frequency domain. This is because any noise contains both stationary and non-stationary parts, while spectral analysis of noise in frequency domain is adequate only if the noise is stationary. A non-stationary part of noise affects observed values of pulsar's parameters, increases their variances, gives rise to larger correlations and makes some of the parameters biased [\cite{1999MNRAS.305..563K}].

Another problem relates to the ergodicity of the overall timing process. There is only one, available to us, realization of the pulsar timing series. It means that one can not access the statistical ensemble to process the data. Instead, we have to apply time integration to evaluate the parameters. If the ergodicity is violated the true value of the parameters will be biased, thus, affecting our ability to maintain the pulsar time scale and to test general relativity.

Third problem is related to the mathematical structure of the estimators of pulsar's parameters. The estimators are integrals in the frequency domain with a kernel being the spectrum of the noise. Spectrum of the {\it red} noise is divergent, and so are the estimators. We do not know what is the best mathematical technique for making the red-noise estimators unbiased and convergent. Definite progress in this area was achieved within the theory of distributions [\cite{2004MNRAS.355..395K}].

Continuous monitoring of pulsar rotational/orbital phase is not sufficient for establishing a pulsar time scale -- an ensemble of pulsars, called a pulsar timing array (PTA), is necessary [\cite{1990ApJ...361..300F}]. Currently, two PTAs are continuously monitored by Australian [\cite{2008AIPC..983..584M}] and Russian [\cite{2006ChJAS...6b.148I}] pulsar groups. A PTA can be also used as a gravitational-wave telescope that is sensitive to radiation at nanohertz frequencies, and for improving the planetary ephemeris [\cite{2009IAU...261.1103H}].

\begin{acknowledgments}
I acknowledge the 2009 summer fellowship support of the Research Council of the University of Missouri-Columbia.
\end{acknowledgments}

\end{document}